\documentclass[a4paper,twoside]{article}
\newcommand{\affil}[1]{$^{\rm #1}$}
\usepackage[authoryear]{natbib}
\bibpunct{(}{)}{;}{a}{}{,}
\usepackage{graphicx}
\date{} 
\title{\large\bf\flushleft Centaurus A at Ultra-High Energies}
\author{\parbox{\textwidth}{\flushleft
\vspace{-0.5cm}
{\it 
Roger W.\ Clay\affil{A,C}, Benjamin J.\ Whelan\affil{A},
and Philip G.\ Edwards\affil{B}}\\
\vspace{0.4cm}
{\small \affil{A}\,School of Chemistry and Physics, University of Adelaide, Adelaide SA 5005}\\
{\small \affil{B}\,CSIRO ATNF, Narrabri Observatory, Locked Bag 194, Narrabri NSW 2390}\\
{\small \affil{C}\,email: roger.clay@adelaide.edu.au}}}
\begin{document}
\twocolumn[
\begin{changemargin}{.8cm}{.5cm}
\begin{minipage}{.9\textwidth}
\vspace{-1cm}
\maketitle
%
%
\small{\bf Abstract: We review the importance of Centaurus~A in
high energy astrophysics as a nearby object with many of the
properties expected of a major source of very high energy cosmic
rays and gamma-rays.  We examine observational techniques and the
results so far obtained in the energy range from 200\,GeV to above
100\,EeV and attempt to fit those data with expectations of
Centaurus~A as an astrophysical source from VHE to UHE energies.  }

\medskip{\bf Keywords:} acceleration of particles ---  galaxies: active ---  gamma rays: observations --- cosmic rays

\medskip
\medskip
\end{minipage}
\end{changemargin}
]
\small

\section{Introduction}

The field of very high energy astrophysics deals with processes
associated with the acceleration and interactions of particles at
energies above those accessible with spacecraft observatories,
characteristically above a few 100\,GeV, up to the highest particle
energies found in nature, above 100\,EeV.  The massive particles at
these energies are known as cosmic rays and at the top of the
energy range are referred to as ultra high energy (UHE).  The
observed energetic photons, which are seen at lower energies, are
known as very high energy (VHE) gamma-rays.

The all sky cosmic ray spectrum exhibits a very steep dependence of flux
against energy.  It extends over 30 orders of magnitude of flux
and ten orders of magnitude in energy to above 100\,EeV with
rather little deviation from a featureless power law
relationship.  There is a steepening at PeV energies, known as
the ``knee'' \citep{HillasARAA}.  At energies in the EeV range,
there is then a flattening known as the ``ankle''.  The knee is
thought to represent either an energy limit to the acceleration
ability of most galactic sources or a limit to the ability of our
Milky Way galaxy to securely contain and build up an internal
cosmic ray flux.  The ankle is thought to represent a change from
predominantly galactic sourced cosmic rays to an extragalactic
flux \citep{GaissStan}.  It is thus most reasonable to look at
energies above a few EeV for direct observational evidence of
Centaurus~A (Cen~A) as a cosmic ray source.

The origins of cosmic rays are not securely known.  It is thought
that supernovae or supernova remnants (SNR) are the most probable
origins of cosmic rays which originate in the Milky Way and that
such particles are energised through diffusive shock acceleration
\citep{ProthClay}.  There appear to be severe limitations to
energies accessible through this process and the highest energy
cosmic rays are postulated to originate in some different
environment outside our galaxy \citep{HillasARAA}.  Cen~A, our
closest active galaxy, is a relatively local extragalactic object
which may contain regions such as its extended radio lobes, or
supermassive central black hole, with physical properties which
enable cosmic ray acceleration to exceed energy limitations which
apply in galaxies like the Milky Way.  For this reason, over
almost 40 years, Cen~A has been the target of observational
searches for evidence that it is a significant VHE or UHE source.
We briefly review the techniques used to study Cen~A, review the
reasons why Cen~A is an attractive observational target, and
examine the observational progress which has been made.

\section{Particle Acceleration \\ Processes and Sites}

There is a general expectation that cosmic ray particles
primarily receive energy through diffusive shock acceleration
\citep{Berezinskii,ProthClay}.  This is a process whereby charged
particles diffusively cross an astrophysical shock front multiple
times, receiving a boost in energy with each crossing.  This
concept has been shown to have the happy characteristic that a
power law cosmic ray energy spectrum results.  Such acceleration
processes were first proposed by Fermi who noted that head on
collisions with moving magnetic clouds resulted in a transfer of
energy to already energetic particles, and that head on
collisions were statistically preferred over others
\citep{Berezinskii}.  Fermi's original process proved to be very
slow and the realisation that multiple (statistical) crossings of
a shock front provided a much faster (``first order'') process
appeared to provide a practical acceleration model.  In such a
picture, there is a clear requirement that (a subset of) the
accelerating particles are repeatedly scattered across the shock.
An energy upper limit of the process results from properties of
the source region which finally fail to provide sufficient
scattering at the highest energies.  

An alternative non-stochastic scenario is that acceleration is
associated with the voltage drop created by a rapidly
spinning supermassive black hole threaded by magnetic fields
induced by currents flowing in a surrounding disk or torus
\citep{Levinson}.  In this case, the maximum achievable energy is
apparently in the UHE region although more detailed modelling
will be required to clearly determine limits imposed by energy
loss mechanisms such as curvature radiation.

Cen~A contains a supermassive black hole and also exhibits evidence of
substantial shocks with evidence for particle
acceleration associated with their related jets
\citep{Hardcastle}.  We shall see below that it has regions
which are capable of scattering particles magnetically as
required.  Whether those necessary conditions are sufficient for
the acceleration of particles to ultra high energies is
the question to be answered observationally.

\section{Particle Propagation and Attenuation}

At very high energies, astrophysical particles are capable of
having inelastic and elastic collisions with particles and fields
in the source, and between the source and our observatories
within the Milky Way galaxy.  These interactions are important to
understanding the astrophysics of the particles which are
observed.  Although limited in size, source regions can contain
strong magnetic fields, intense photon fields over a great energy
range, and a high plasma density.  In intergalactic space, our
knowledge of fields and energy densities is limited but we
expect, at least, that there will be some magnetic fields,
starlight, infra-red radiation, and the cosmic microwave
background \citep{Driver}.  Closer to home, messenger particles
will transit whatever fields are contained in our local group of
galaxies, our galactic halo and the plane of the Milky Way.  In
the latter context, it is expected that the known magnetic fields
of our galaxy (with an underlying regular field at levels of a
few $\mu$G but up to 10\,$\mu$G if a random component is included
\citep{Sun}) will have deflected all charged cosmic rays by
significant amounts \citep{Stanev}.  Astrophysical angular
uncertainties then exceed instrumental ones for charged cosmic
rays.

It is possible for accelerated protons to interact (most likely
in a source region) and convert to neutrons.  This, for instance,
is an argument for a possible cosmic ray excess in the direction
of our galactic centre since the neutrons will not suffer
deflection in galactic plane magnetic fields which would
certainly scatter protons out of a directional beam
\citep{ClayGC}.  Isolated neutrons decay within a few minutes
when at rest but cosmic ray neutrons with a high
relativistic gamma factor will survive long distances in the
laboratory frame.  However, at the distance of Cen~A, neutrons
with energies below 400\,EeV (just above the highest energy
cosmic ray recorded from any direction \citep{BirdFE}) would decay before
reaching us.  We note that such decay is statistical and some
neutrons might be observable even at lower energies but the
resulting flux would be greatly attenuated below 100\,EeV.  It
seems that neutrons generated within a Cen~A source will not
provide a direct undeflected beam, although protons could
interact in some intermediate matter, resulting in a ``halo''
around the direction of a source.

Cosmic ray interactions can additionally produce a flux of high
energy neutrinos.  These could be through interactions with the
CMB or with particles and fields close to the source.  In the
latter case, LUNASKA \citep{Lunaska} or northern UHE neutrino
detectors such as ANTARES \citep{ANTARES} might search for signals from the
direction of Cen~A.

\section{Interactions with Photon \\Fields}

Our Universe is known to be pervaded by the cosmic microwave
background (CMB) having a photon number density a thousand times
that of characteristic plasma densities.  Despite the low
energies of microwave photons, VHE photons and UHE nuclei see
them as significant targets over modest astrophysical distances.
Photons from Cen~A are expected to be severely attenuated
\citep{ProthMNRAS, ProthRev} over a range of energies, with a
spectral cut-off beginning in the energy range 130 to 200\,TeV
depending on the strength of the intergalactic magnetic field
\citep{BP3}.  In this absorption feature, attenuation lengths of
a few kpc are expected for its deepest point at a little over
1\,PeV.  The absorption feature progressively weakens at higher
energies and, for sources within our galaxy, at much higher
energies the absorption dip may be passed and photon attenuation
may be reduced \citep{ProthMNRAS}.  However, for photons from the
more distant Cen~A with a much greater path length, the
absorption will be strong up to 10\,EeV \citep{ProthRev}.  The
Pierre Auger Observatory (see Section 11) is capable of selecting
photons from its overall detected flux at such energies and a
search for photons from Cen~A above 10\,EeV would seem
worthwhile.

Cosmic ray nuclei will also interact with the CMB but this is not
important until much higher energies than for photons, and the
attenuation length is much greater.  This attenuation phenomenon
is conventionally named the GZK effect after the people 
(Greisen, Zatsepin and Kuzmin) who
proposed it in 1966 for cosmic ray protons interacting on the CMB
\citep{Berezinskii}.  The interaction has a proton energy
threshold of about 60\,EeV, a factor of 100,000 times greater than
energies associated with the photon attenuation.  There are also
interaction processes for other nuclei on the CMB which become
important at about this energy.  However, whilst the
characteristic attenuation length due to the GZK effect is of the
order of 100\,Mpc, the interaction mean free path is of the order
of the distance to Cen~A.  This is due to the modest energy
loss per interaction.  Thus both VHE gamma-rays and UHE cosmic
rays sourced from Cen~A are expected to show evidence of
interactions with the CMB.

It appears that attenuation compatible with the GZK effect is
evident in the all sky cosmic ray spectrum of the Pierre Auger
Observatory \citep{PAOGZK}, although evidence for a propagation
cut-off requires an assumption that the cosmic ray source
spectrum does not contain a similar feature.  The current dataset
at these energies is small and it is questionable if any
presently observed events from the direction of Cen~A could show
a statistically convincing evidence for the existence of, or a
lack of, GZK attenuation (although see Section 12).

\section{Magnetic Fields}

Charged cosmic ray particles will have their propagation
directions changed in their passage through astrophysical
magnetic fields.  That deflection will depend on the particle
rigidity, the ratio of momentum and charge.  At the energies of
interest here, this is effectively the ratio of the energy and
the charge.  A convenient rule of thumb is that the radius of
curvature of a 1\,PeV proton trajectory perpendicular to a
uniform 1 microgauss magnetic field is 1\,pc.  Characteristic
galactic fields are at these levels or just above, but their
structure and, particularly, their extension out of the galactic
plane are poorly known.  Also, their strength within our local
group of galaxies and the remaining intergalactic space between
ourselves and Cen~A is largely unconstrained by observation
\citep{Beck}.  There is evidence that richer groups may be
pervaded by multiple $\mu$G level magnetic fields
\citep{Clarke,HollittMag} but this may not apply in our local
region.  This lack of knowledge is a major problem since, even at
proton energies of 100\,EeV, we are still dealing with a radius
of curvature of only 100\,kpc in a characteristic magnetic field.
As a result, we are unable to specify whether charged cosmic ray
propagation over a distance of 3.8\,Mpc from Cen~A is diffusive
(with an uncertain magnetic turbulence scale size) or whether we
can assume roughly linear propagation.  A clear scattered cosmic
ray signal from Cen~A would give us invaluable information
regarding our local extragalactic magnetic fields (though of
course this requires a knowledge of the intrinsic source size).

\section{The Air Shower Technique }

Our atmosphere is opaque to primary radiation at energies with
which very high energy astrophysics deals.  Also, the flux of
astronomical particles becomes sufficiently low at those energies
and above such that direct satellite observation ceases to be
effective for reasonable spacecraft detector collecting areas.
The effective way of working at higher energies is through the
use of our atmosphere as a target and observing the cascades of
particles, known as ``air showers'' or ``extensive air showers
(EAS)'', which are produced as incident particles deposit their
energy, first as conversion to secondary particle mass and
kinetic energy and then into atmospheric gas excitation and
ionisation.  A good introduction to the physical processes in air
showers can be found in \citet{Allan}.

Primary gamma-rays initiate cascades which, to a first
approximation, develop through successive processes of pair
production and then bremsstrahlung of the daughter electrons and
positrons.  The mean free paths for those processes are related
and are between 30 and 40\,g\,cm$^{-2}$ (compared to 
the thickness of the vertical
atmosphere of about 1000\,g\,cm$^{-2}$).  This rather simple
cascade develops ``exponentially'' in particle \\ number until
ionisation energy losses begin to inhibit further development.
The cascade thus reaches a maximum particle number (often
stated as a number of ``electrons'', $N_e$).  Such cascades are
statistical in character but the short interaction mean free
paths compared to the total atmospheric depth result in a rather
smooth development profile.

Primary nuclei also initiate cascades but these are more complex
and irregularly structured.  They begin with a strong interaction
which produces pions.  The neutral pions decay to a pair of
gamma-rays which then cascade as we have just seen.  However, the
charged pions are likely to decay (they may interact again if the
atmospheric conditions are conducive) to muons.  Those muons will
most likely continue to traverse the atmosphere without further
major interactions, just suffering a continuous energy loss from
their ionisation and excitation of atmospheric gases.  This
cascade now has three components.  They are: the remnants of the
original particle which only loses a fraction of its initial
energy at each interaction (the nuclear core), the muons, and the
electromagnetic cascades.  A key point is that the nuclear core
continues to inject energy through further interactions,
resulting in the initiation of superimposed electromagnetic
cascades.  The overall ``shower'' particle number is then a
superposition of electron numbers in successive electromagnetic
cascades, building and decaying, plus the integrated numbers of
muons.  This picture is further complicated by the fact that the
cosmic ray beam (at least at the lower energies) is a mixture of
nuclear components \citep{GaissStan}.  These various nuclei have their own
interaction mean free paths for initiating cascades (longest for
protons --- 80\,g\,cm$^{-2}$ --- and shorter for more massive
nuclei) and, though difficult, and probably not possible on an
event by event basis, this offers a means for studying the beam
composition, or its change with energy.

All cascades contain charged particles which scatter through
interacting with atmospheric gas.  As a result, the
electromagnetic cascades spread laterally with a characteristic
distance of below 100\,m for the numerically dominant
electromagnetic component.  However, some electromagnetic
particles can scatter to very large ``core distances'' and
ground-based detecting arrays such as the Pierre Auger
Observatory record particles at kilometres from
a lateral extension of the original cosmic ray trajectory.  The muons
scatter rather little but retain their direction from their
initiating interaction which results in a characteristic spread
of hundreds of metres at sea level.  Again, a very few can be
found kilometres from the core.

Air shower cascades are studied by sampling a selection of their
components.  This is efficient in terms of enabling the detection
of rare events (at the highest energies the flux may be measured
in terms of km$^{-2}$\,century$^{-1}$).  This sampling can be
accomplished with a sparse array of ground-based charged particle
detectors which sample the cascade at a single development level.
The requirement that particles reach the ground limits this
technique to energies above about 100\,TeV at sea level or to
detector arrays located at very high altitudes \citep{Tibet}.
Alternatively, that observational energy threshold can be reduced
by detecting the bright beam of forward directed \v{C}erenkov
light produced in the atmosphere using large optical photon
collecting telescopes which ``image'' those photons onto a
photomultiplier ``camera''.  The VHE gamma-ray telescopes such as
H.E.S.S.\ \citep{HESS} fall into this category.  At the highest
energies, where a low level of light emission per shower particle
is not a limiting factor, isotropic nitrogen fluorescence light,
produced by the cascade exciting atmospheric gas, can be
successfully recorded.  This enables a large collection area to
be achieved with large mirrors viewing the cascade from the side.
The Pierre Auger Observatory employs this technique together with
a large array of ground-based large area particle detectors.

\section{Differentiating between \\Gamma-Rays and Cosmic \\Rays}

In recent years, the study of very high energy gamma-rays has
become an important component of astrophysics.  This has become
possible partly through improvements in instrumental sensitivity
and angular resolution but, also, through significant improvements
in the software discrimination between gamma-ray induced
extensive air showers and the numerically dominant cosmic ray
showers. The gamma-ray showers are predominantly electromagnetic
with interaction processes (pair production and bremsstrahlung)
which occur at rather small intervals in the atmosphere.  The
resulting cascades are rather simple and smooth.  On the other
hand, cosmic rays initiate and feed cascades through a nuclear
process with a much longer mean free path and their interactions
produce muons in addition to electromagnetic particles.  The
cascade development is then rather irregular and also has an
irregular geometrical spread due to the muons, which can travel,
with rather little scattering, at significant angles away from
the central cascade core.  Differentiation between gamma-ray and
cosmic ray initiated cascades conventionally depends on vetoing
cosmic ray cascades.  This is achieved either by detecting an
irregular shower development (or irregular \v{C}erenkov image), a
development which peaks at an atmospheric depth characteristic of
nuclei for a given total energy (or particle content), or a muon
content greater than expected for a gamma-ray cascade.

VHE gamma-ray astronomy using the atmospheric \v{C}erenkov
technique has proved to be very efficient in producing images
which have good gamma-ray to cosmic ray discrimination.  This is
due to careful Monte Carlo modelling of the cascade and imaging
processes to develop suitable image cuts \citep{HESS}. These are
broadly based on vetoing the larger, less constrained, cosmic ray
images, together, in some cases, with a requirement 
that point sources under study
are at known positions in the image.  

PeV gamma-ray studies have more commonly explicitly used a muon
veto in which cascades with significant muon numbers have been
rejected.  This approach has had mixed success.  At even higher
energies in the EeV range, gamma-ray initiated showers are
expected to reach maximum development deeper in the atmosphere
than cosmic ray showers and work is ongoing to select potential
gamma-ray cascades on this basis.  Presently, the Pierre Auger
Observatory claims upper limits to the UHE photon fraction using
this method \citep{PAOphoton}.

\section{Searches at TeV Energies}

The first searches for VHE gamma-ray emission from Cen~A were
made with the Narrabri Stellar Intensity Interferometer.  The
Intensity Interferometer consisted of two 6.5\,m diameter
segmented optical reflectors mounted on a 188\,m diameter
circular track.  The interferometer took advantage of the
Bose-Einstein statistical nature of light, with optical photons
from stars tending to arrive in clumps.  The correlations in
intensity between the two reflectors enabled the diameters of
bright stars to be inferred. The light pool from \v{C}erenkov
radiation produced by VHE cosmic rays has a similar extent to the
track diameter, and so calculations and tests were performed to
confirm that the \v{C}erenkov light signal was not contaminating
the measurements of steller diameters \citep{HBDA}. It was
recognised, however, that the Intensity Interferometer could also
be put into service as a \v{C}erenkov detector.
\citet{Grindlay2} used a 120\,m separation between reflectors and
operated the two as a stereo detector. They also employed a novel
background rejection scheme, using off-axis photomultipliers to
detect \v{C}erenkov light from the penetrating muon component of
cosmic-ray initiated cascades, which allowed $\sim$30\% of
recorded events to be rejected. Between 1972 and 1974, a sample
of 11 sources, including pulsars, X-ray binaries, the Galactic
Centre, and AGN, were observed, with source selection based on
X-ray and SAS-2 (30\,MeV to 100\,MeV) gamma-ray results.  A
time-averaged 4.5$\sigma$ excess was detected in a total
observing time of 51\,hr on Cen~A \citep{Grindlay1},
corresponding to a integral flux above 300\,GeV of
(4.4$\pm$1)$\times$10$^{-11}$\,cm$^{-2}$\,s$^{-1}$.  As the outer
radio lobes were well outside the beam, these were excluded as
the source. Possible variability of the gamma-ray signal
\citep{Grindlay1}, coupled with theoretical modelling, also
excluded the inner radio lobes as the gamma-ray source, with a
proposed model of the gamma-ray flux arising from inverse Compton
scattering in the nucleus of Cen~A being favoured
\citep{Grindlay3}.

Subsequent VHE observations
over the next 35 years yielded negative results.  The Durham
group, based at Narrabri, observed between March 1987 and April
1988 with their MkIII telescope for a total of 44 hr of good
on-source data.  The 3$\sigma$ flux upper limit above 300\,GeV
was 7.8$\times$10$^{-11}$\,cm$^{-2}$\,s$^{-1}$ \citep{Durham1}.
In March 1997, 6.75 hours of observations were made with the
Durham Mk6 telescope, which provided better discrimination
against the cosmic ray background, with a 3$\sigma$ flux upper
limit above 300\,GeV of
\\5.2$\times$10$^{-11}$\,cm$^{-2}$\,s$^{-1}$ \citep{Durham2}.

The JANZOS group observed Cen~A from New Zealand for 56.9 hr between 
April 1988 and June 1989, reporting a 95\% confidence level upper 
limit on the flux above 1\,TeV of
2.2$\times$10$^{-11}$\,cm$^{-2}$\,s$^{-1}$ \citep{JANZOSTeV}.

Interest in VHE emission from Cen~A was rekindled by the EGRET
detection of Cen~A in the 30\,MeV to 30\,GeV range
\citep{Steinle98,Sreekumar}.  The CANGAROO 3.8\,m telescope was
used in March and April 1999 to record a total of 45\,hr of on- and
off-source data. The resulting 3$\sigma$ flux upper limits above
1.5\,TeV were 5.5$\times$10$^{-12}$\,cm$^{-2}$\,s$^{-1}$ for a
point source at the core of the galaxy, and
1.3$\times$10$^{-11}$\,cm$^{-2}$\,s$^{-1}$ for an extended
region of radius 14$'$ centred on the core \citep{CANGAROO1}.  In
March and April 2004, further observations were made with three
10\,m telescopes of the CANGAROO-III array. From 10.6 hours of
on-source data, upper limits were set for the several regions of
interest: for the core of Cen~A the 2$\sigma$ upper limit above
424\,GeV was 4.9$\times$10$^{-12}$\,cm$^{-2}$\,s$^{-1}$
\citep{CANGAROO2}.

The H.E.S.S.\ group achieved a detection of Cen~A with over
120\,hr of observation between April 2004 and July 2008. Their
measured integral flux above 250\,GeV was
(1.56$\pm$0.67)$\times$10$^{-12}$\,cm$^{-2}$\,s$^{-1}$.  The
detection is concentrated on the central galaxy region, not the
lobes.  This excess ``only matches the position of the core, the
pc/kpc inner jets and the inner radio lobes'' \citep{HESScena}.

The H.E.S.S.\ flux is a factor of almost 30 below the original report
of \citet{Grindlay1}.  However, the Intensity
Interferometer observations were made during an extended period of
enhanced X-ray emission in the early 1970s. Although X-ray monitoring
was infrequent during the 1980s, it appears Cen~A has been in a
relatively quiescent state for most of the last 30 years
\citep{Bond,Turner,Steinle06}. A low X-ray state would 
plausibly result in a low flux of inverse Compton scattered VHE gamma-rays.

TeV detections and upper limits are plotted in Figure~1.

\section{Searches at PeV Energies}

Searches at PeV energies are made using air shower arrays of
particle detectors. The Buckland Park air shower array was used
between 1978 and 1981 to study the anisotropy of cosmic rays
above an energy of 1\,PeV. The directional accuracy of the array
was 3$^\circ \times$\,sec($\theta$), where $\theta$ is the zenith
angle. The study confirmed that the overall cosmic ray flux shows
no strong sidereal isotropy at these energies, and the work was
then extended to search for more localized excesses. The least
isotropic declination band was that between $-40^\circ$ and
$-45^\circ$, and the largest excess in bins of 1\,hr in right
ascension coincided with Cen~A. The overall significance was
2.7$\sigma$, a value not unexpected by chance given the number of
bins examined, but which encouraged further investigation.  There
was some evidence in the binned data for excess event numbers
toward both outer radio lobes \citep{BP1}.  Some supporting evidence was also
noted in two other Southern Hemisphere experiments \citep{FS54,
Kamata}, though with differing energy thresholds, angular
resolutions and years of operation. It was also acknowledged
that, at PeV energies, the signal could not be due to gamma-rays
from Cen~A as the path length for interactions of PeV gamma-rays
with CMB photons is only 10\,kpc \citep{BP1}.

The JANZOS experiment combined TeV telescopes with a PeV
scintillator array, and a search for gamma-rays above 100\,TeV
was conducted with data taken between October 1987 and January
1992. No significant excess was found over this complete time
range, but an excess was observed over 48 days from April to June
1990, with the excess concentrated in events with enegies below
200\,TeV, consistent with the effects of the expected absorption
at higher energies. Monte Carlo simulations were used to derive a
probability of 2\% for the observed 3.8$\sigma$ excess to arise
by chance \citep{JANZOSPeV}. A contour map of the significance
showed a peak that coincided, within the angular resolution of
the array, with the core of Cen~A.
Buckland Park data taken between March 1988
and February 1989 was examined and no significant excess found
from Cen~A \citep{BP2} --- this being consistent with the JANZOS
result for this period.

A larger Buckland Park data-set, from 1984 to 1989, was split,
{\it a priori}, into three event size bins, and a excess found in
the lowest size bin, corresponding to energies below 150\,TeV
\citep{BP3} (hence below the CMB absorption feature).  A
confidence level of 99.4\% was claimed for the excess.  A
Kolmogorov-Smirnov test indicated there was no significant
evidence for enhanced periods of emission over this five-year
period.  A contour map of the excess showed a peak suggestive of
a point source compatible with the core of Cen~A.

PeV detections and upper limits are also plotted in Figure~1.

\begin{figure}[h]
\begin{center}
\includegraphics[scale=0.32, angle=-90]{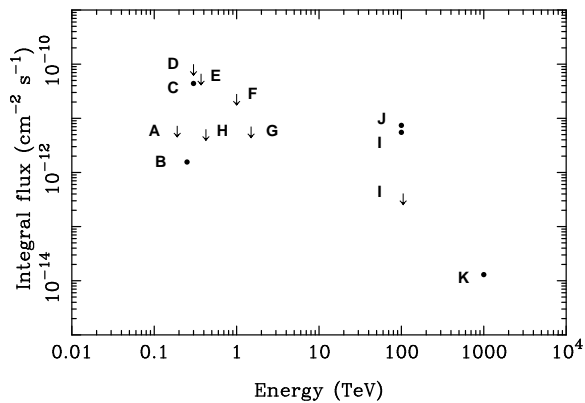}
\caption{Reported detections (solid circles) and upper limits (arrows)
to the flux from Cen~A at TeV and PeV energies.
(The labels are as follows:
A, H.E.S.S., Aharonian et al. 2005;
B, H.E.S.S., Aharonian et al. 2009;
C, Narrabri, Grindlay et al. 1975a;
D, Durham, Carraminana et al. 1990
E, Durham, Chadwick et al. 1999;
F, JANZOS, Allen et al. 1993a;
G, CANGAROO, Rowell et al. 1999;
H, CANGAROO-III, Kabuki et al. 2007;
I, JANZOS, Allen et al. 1993b;
J, Buckland Park, Clay et al. 1994;
K, Buckland Park, Clay et al. 1984.)
The Chadwick et al. 1999 point (E) has been moved horizontally from the
actual 300\,GeV
energy threshold for clarity.
}\label{fig1}
\end{center}
\end{figure}

\section{SUGAR}

The Sydney University Giant Air Shower Recorder (SUGAR)
\citep{SUGARarray}, apart from having a creative acronym, was
notable and important in pioneering the start of a new era of
cosmic ray study.  Like the huge Pierre Auger Observatory
(below), with an enclosed area of 70\,km$^2$ SUGAR's design
recognised that the flux of cosmic rays at the highest energies
is so low that its ground detectors required large separations
and could not realistically be connected by cable to their direct
neighbours.  They required a measure of local autonomy.  In the
case of SUGAR, this was through the realisation of a local
coincidence between two muon detectors at each station and the
tape recording of their data plus time stamping for later global
array analysis.  The detector sites were also autonomous in terms
of power, generating their own power thermoelectrically.
Previous arrays had detected real time coincidences between
spaced detectors to initiate data acquisition following the
arrival of a suitably energetic shower.  SUGAR operated for long
enough to detect a significant number of showers with energies
above the GZK cut-off energy and, until the commissioning of the
Pierre Auger Observatory, was the only array at the highest
energies with Cen~A in its field of view.  SUGAR was shown to
have a problem with photomultiplier afterpulsing which could make
some energy assignments uncertain, although its direction
determinations would not have been affected by that.  Data from
this array was used for studying a possible association of the
highest energy cosmic rays with the direction of our galactic
centre \citep{SUGARGC}.  No excess from the direction of Cen~A was
evident in that analysis.
Figure~2 shows the SUGAR highest energy
events \citep{SUGcat} in the vicinity of Cen~A (for comparison
with Figure~3 for the Pierre Auger Observatory).

\begin{figure}[h]
\begin{center}
\includegraphics[scale=0.4, angle=0]{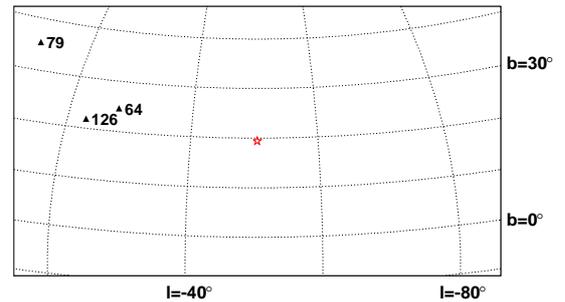}
\caption{SUGAR events with energies above 60\,EeV in the vicinity
of Centaurus~A \citep{SUGcat}.  The events are labelled by their
energy in EeV.  The red star indicates the position of
Cen~A.}\label{fig2}
\end{center}
\end{figure}

\section{The Pierre Auger Observatory}

The Pierre Auger Observatory (PAO) is the largest cosmic ray
detector ever built. It currently consists of a Southern
Hemisphere site located near the town of Malarg\"{u}e, Argentina,
with planning of a Northern Site in Colorado, USA,
underway \citep{Harton}. The southern site covers an area of
approximately 3000\,km$^{2}$ \citep{Suom}. The PAO
employs two cosmic ray detection methods through the Surface
Detector (SD) and the Fluorescence Detector
(FD) \citep{Proto,Bellido,SD}.

The SD consists of more than 1600 autonomous particle detector
stations which employ the water-\\\v{C}erenkov detection technique.
The particle detectors operate by recording \v{C}erenkov light
emitted when relativistic charged particles in an air shower pass
through 1.2\,m deep purified water enclosed in large-area (10\,m$^2$)
tanks. These stations are arranged on a triangular grid, with
1.5\,km spacing. This spacing results in the SD being fully
efficient for detecting showers with a primary energy of above
$3\times 10^{18}$\,EeV at zenith angles of 60$^{\circ}$ or less
\citep{Suom}. Statistical energy uncertainties from the SD are
approximately 17\%, with an additional systematic uncertainty of
7\% at $10^{19}$\,eV (increasing to 15\% at $10^{20}$\,eV)
arising from calibration with FD energies (see below)
\citep{Guilio}. Directional uncertainties are approximately
$1.5^{\circ}$ at energies around 3\thinspace EeV, reducing to
less than $1^{\circ}$ for energies above approximately 10\,EeV
\citep{Bonifazi}. It has a duty cycle of slightly less than 100\%
giving a current integrated exposure of more than
12,000\,km$^{2}$\,sr\,yr, increasing by approximately
350\,km$^{2}$\,sr\,yr per month \citep{Schuss}.

The FD makes use of the air fluorescence method. 
Twenty-four telescopes
are separated into 4 groups of 6 telescopes (each group being
termed a FD `site'), which overlook different sections
of the SD. Each telescope views $30^{\circ}$ in azimuth, giving
each site a 180$^{\circ}$ azimuthal field of view, and 28.6$^{\circ}$ in
elevation. In each telescope a camera consisting of an array of
photomultiplier tubes, viewing separate regions of sky,
collects the light emitted by nitrogen molecules excited by the
EAS. Using pulse timing information from triggered pixels, the axis
along which the shower front propagates can be reconstructed. The
energy of the shower, apart from a small amount of `invisible
energy' carried by neutrinos and muons which are not visible to
the FD, is proportional to the integrated light flux along the
shower's path. This allows an effectively calorimetric
measurement of particle energy \citep{Bellido}. The statistical
energy uncertainties are approximately 9\%, with a systematic
uncertainty of approximately 22\% arising from factors such as
limitations of knowledge of the instantaneous atmospheric
profile and aerosol content \citep{Guilio}. The
requirement of clear, moonless nights for the operation of the FD
means that its duty cycle is about 13\% \citep{Schuss}.

The colocation of the FD and SD allows events to be observed by
both detectors. The events for which this occurs are termed
`hybrids'. For these events, the timing information from a
triggered SD station is added to that from the FD trace to
reconstruct the arrival direction \citep{Bellido}. This allows a
greater accuracy in determining the shower axis than is possible
with either method alone, and the average hybrid directional
uncertainty is 0.6$^{\circ}$ \citep{Bonifazi}. An additional
advantage of the hybrid method is that it allows the SD energy
scale to be determined. By a comparison of independent SD and FD
reconstructions of the same events, SD energies can be calibrated
against the essentially calorimetric values from the FD
\citep{Guilio}.

Composition studies are performed primarily with the FD through
measurements of the position of shower maximum, $X_{max}$. This
value indicates the slant depth in the atmosphere, in
g\,cm$^{-2}$, at which the flux of fluorescent light from the EAS
reaches its maximum. From shower to shower the value of $X_{max}$
fluctuates due to the statistical nature of shower initiation and
development. On average, however, nuclei are expected to have
smaller $X_{max}$ values than protons at a given energy, and
fluctuations in $X_{max}$ are expected to be
smaller. Consequently, FD measurements are used to study the
behaviour, as a function of energy, of both $\langle
X_{max}\rangle$ (the energy dependence of which is termed the
`elongation rate') and the RMS of $X_{max}$ to look for possible
changes in the composition of the primary particles
\citep{Bellido2}. It should be noted, however, that the
interpretation of these results is not clear due to uncertainties
in hadronic interaction physics at such high energies
\citep{Bellido2,Durso}.

The large duty cycle of the SD makes the prospect of utilizing it
to determine CR composition highly desirable. This is a somewhat
more difficult task than with the FD, however, as a direct
measurement of $X_{max}$ is not possible with the SD. Methods
such as studying the risetime of the particle signal in the SD stations,
the shower front radius of curvature, the ratio of the muonic to
electromagnetic contributions to the signal and the azimuthal
asymmetry in station signal around the shower axis are currently
being investigated for their suitability in composition
determination with the SD \citep{Wahlberg}.

\section{Observational Results at the Highest Energies}

The Pierre Auger Observatory is the only system presently
recording data at EeV energies from the direction of Centaurus~A.
It presented its first skymap in 2007 \citep{PAOScience}
displaying the 27 highest energy events at that time.  This map
appears to show event clustering in the general direction of
Cen~A.  Data in that skymap which are in the vicinity of Cen~A
are shown in Figure~3 \citep{PAOlong} which illustrates apparent
clustering around the direction of Cen~A.

\begin{figure}[h]
\begin{center}
\includegraphics[scale=0.4, angle=0]{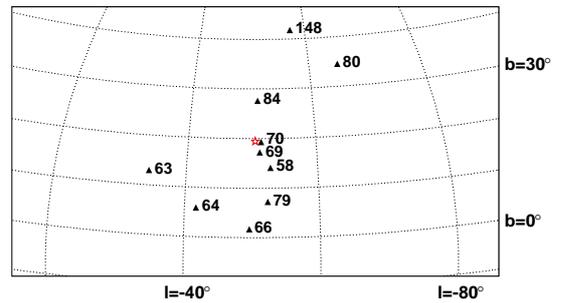}
\caption{Auger events above 57\,EeV in the vicinity of Centaurus~A
\citep{PAOlong}.  The events are labelled by their energy in
EeV.}\label{fig3}
\end{center}
\end{figure}

The Auger paper interprets the dataset as a whole as
being statistically associated with the directions of local AGNs.
This was on the basis of an {\it a priori} search
``prescription''.  The Pierre Auger Collaboration has not yet
developed a discovery prescription for Cen~A and, as a
result, no {\it a priori} statistical analysis is presently
possible.  However, one can comment on the properties of the
dataset.

\citet{HillasAuger} has independently examined this Auger data
set and confirms the conclusion that the highest energy events
are associated with rather typical Seyfert galaxies in clusters
at distances of typically $\sim$50\,Mpc. The clustering of events
near Cen~A is confirmed and an origin in Cen~A, or alternatively,
NGC\,5090 considered.  The close proximity of Cen~A, however, led
Hillas to conclude that Cen~A is probably an inactive
10$^{20}$\,eV accelerator as more distant galaxies play a larger
role than might have been expected on the basis of a simple
inverse square law flux dependence.

More recently \citep{PAOcenaLodz}, the Pierre Auger Collaboration
has shown continuing evidence for a concentration of the highest
energy events in the direction of Cen~A.  Those data show an
excess of events above 55\,EeV within $18^\circ$ of Cen~A which
is well above the 68\% confidence interval for a sample from an
isotropic distribution.  In that range, 12 events are found where
2.7 are expected on the basis of an isotropic flux.
Approximately 30\% of the Auger events show some evidence of
being members of such clustering out to $30^\circ$ from Cen~A,
approximately the same fraction (10/27) as found in the original
Auger skymap.

On the basis of this evidence, one might speculate that Cen~A is
the source of a substantial fraction of the extragalactic cosmic
rays at energies above the ankle of the cosmic ray energy
spectrum.  In that case, one notes that the angular size of the
cosmic ray ``image'' is appreciably greater than the known
physical dimensions of the astronomical source on the sky.  That
cannot be explained by instrumental errors since, as noted above,
the known angular resolution of the observatory at these energies
is below 1$^\circ$ \citep{Bonifazi} and one naturally invokes
magnetic scattering in intergalactic space or within our galactic
region.  If the scattering occurs in intergalactic space, one
might plausibly assume a 10\,kpc turbulence cell size, leading to
a total scattering deflection of the order of 20 times the
scattering in an individual cell (after the passage of
approximately 400 individual cells).  To fit the observed excess
around Cen~A, this requires a turbulent intergalactic field of
strength 0.1\,$\mu$G.

At these energies, any deflections from the direction of
Cen~A in known regular galactic plane magnetic fields are
likely to be modest (a few degrees \citep{Stanev}) but the extent
and strength of magnetic fields in any galactic halo around the
Milky Way are unknown \citep{Sun} and could plausibly be substantial
if the dimensions of the halo are large.  The Auger excess
appears to limit the possibility of such effects from regular
fields since there seems to be no great asymmetry perpendicular
to the galactic plane although there may be an oval axis in that
direction for the excess just discussed.

In a scenario in which the object Cen~A is the origin of
the Pierre Auger ``Centaurus A'' excess, one must explain the
apparent similarity between the distribution of energies within
the excess when compared with the totality of Auger events.
Using the published event data \citep{PAOlong}, there is no significant
difference between the two spectra (Figure~4) on the basis of a
Kolmogorov Smirnov test (at $>$ 99\% level), admittedly with a very
limited event list.  If that continues to be the case, one might
have to abandon the GZK cut-off as the source of a deficit of
events above 60\,EeV and argue for a source acceleration
limitation.

\begin{figure}[h]
\begin{center}
\includegraphics[scale=0.4, angle=0]{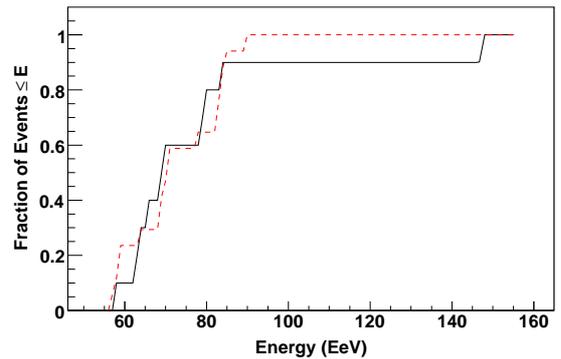}
\caption{Energy distribution for events inside and outside a $25^\circ$
circle centred on Cen~A \citep{PAOlong}. The dotted red line is for the
events more than 25$^\circ$ from Cen~A.}\label{fig4}
\end{center}
\end{figure}

An examination of event data presented by the Pierre Auger
Collaboration with their 2007 sky map of the highest energy
events shows a possible systematic variation in particle energy
across the Cen~A excess, with the highest energy events being at
the highest (positive) Galactic latitudes.  Such an effect could
be due to chance, but it could also count as evidence against the
excess being associated with Cen~A since there is not now a
symmetry in the event energies about the central region.  An
explanation could be that the cosmic ray source for the highest
energy events is in the northern lobe, or there could be
contamination from another source in the supergalactic plane, or
simply that the combination of intergalactic scattering and the
passage through structured regular fields combine to produce the
effect.

The energy variation with direction could also suggest that there
is another possible approach to understanding the ``Centaurus A''
excess.  This is that the propagation is dominated by regular
intergalactic fields and that the particles are deflected as in a
magnetic spectrometer, with the true source being in the
direction from which an ``infinite'' energy particle would have
been seen.  Since the excess has its highest energy particles
furthest north from the galactic plane, one would invoke a
magnetic field parallel to that plane (perpendicular to the
supergalactic plane) with a ``true'' source region some distance
further to the north such that the angular deflection is
inversely proportional to the particle rigidity.  In this
scenario, it could be that the deflecting magnetic field is a
halo field of the Milky Way.  This would require a product of
magnetic field strength for the regular component of the field
and its spatial extent of the order of 50\,$\mu$G\,kpc. This may
not be incompatible with models of magnetic fields in groups of
galaxies or extended galactic halos.

\section{Centaurus~A as a Cosmic Ray Source}

As we saw, it is usual to think of cosmic rays being accelerated
to their observed energy in a rather slow statistical process.
There are alternative possibilities, such as a single
acceleration through a very large potential step or a ``top
down'' model in which an ultra-energetic particle is the result
of the decay of an exotic highly massive initial particle.  If
the process is something like diffusive shock acceleration, the
acceleration site must be such that its scattering fields are
capable of returning accelerating particles many times across a
shock front.  This would appear to require local magnetic fields
with products of strengths and physical dimensions such that a
particle radius of gyration at the highest energies can be
contained within the physical boundaries of the field.  This is
often expressed through one of the ``Hillas diagrams''
\citep{HillasARAA}.

When it comes to considering Cen~A as a source, possible extremes
of the spectrum of sites would be within the modest strength
magnetic fields enclosed in one or other of its giant radio
lobes, or (at the other extreme) within very strong fields close
to the central engine.  Somewhere within a jet, or the southern
shock, could also be candidate sites specific to Cen~A.  The
radius of gyration of a cosmic ray proton (in kpc) is numerically
close to its energy (in units of EeV) divided by the magnetic
field strength (in units of microgauss).  Containment within an
acceleration region will require that the region is significantly
larger than that radius of gyration.  Since it is the particle
rigidity which is relevant, this requirement would be eased in
proportion to the nuclear charge for heavier nuclei.  This would
mean that the acceleration of protons in 200\,kpc lobes of Cen~A
(the whole of a lobe) would require magnetic fields filling a
lobe at microgauss levels in order to accelerate particles to the
measured Auger limit of about 200\,EeV.

Under such conditions (\citep{ProthClay} equation 41), a time of
the order of 100 million years is required for the acceleration
process.  This would seem to be approximately the limit of
possible acceleration both under a $10^8$\,year estimate of AGN
lifetimes and an estimate of microgauss strength fields in the
lobes.  The cosmic ray energy spectrum extends over 30 orders of
magnitude in flux and very few accelerated cosmic rays are
required to reach the highest energies - they are statistical
anomalies. The source magnetic field must be strong enough and
large enough in scale for the highest energy particles to be
capable of one last diffusive scattering across the shock front.

An alternative extreme of the spectrum of possible Cen~A
acceleration sites might be within the most central volume of the
AGN, close to the black hole.  
The majority of TeV and PeV detections
are consistent with an excess concentrated on the
central galaxy region, and not the outer lobes.  
A central region with dimensions
of, say, 10\,pc would require a shock containment field strength
approaching 1\,Gauss.  This would be substantial but not
unreasonable.  A difficulty with such a region would be to
accelerate particles to UHE energies over a substantial period of
time within a dense photon field containing photons with energies
substantially above those of the CMB.  This is because cosmic ray
energy loss interactions would be significant from at least EeV
energies.  This attenuation at energies below the ankle of the
cosmic ray spectrum makes it difficult to see how the Auger Cen~A
spectrum could bear similarity with the conventional spectrum
from other directions.

Cen~A is a radio galaxy with highly extended jets and lobes.  As
we noted, it could be that such lobes play a key part in
accelerating particles to the highest observed energies.
\citet{NagarMat} have discussed the possible role of objects with
that morphology as sources of the Auger highest energy events.
They note that there is a number of such sources (5) in the
general vicinity of Cen~A out of a total of 10 in the ``field of view'' of
the PAO.  They also note that such objects seem to be
statistically closely related to the directions of the highest
energy PAO events.  This proposition seems to be arguable but, if
this is the case, the contribution of Cen~A to the total flux
must be below a level proportional to its radio emission since,
including all such objects, Cen~A dominates the total sum of the
radio fluxes typically by at least an order of magnitude
\citep{NagarMat}.  This is due to its proximity to us, and it
could be that some of the other objects are more effective at
accelerating particles to the highest energies.  
We noted that Cen~A may be a variable source at high energies, 
which may support this idea \citep{HillasAuger}.

Recently, \citet{Rieger} have considered Cen~A as a VHE gamma-ray
and UHE cosmic-ray source, and conclude that advection dominated
accretion disk models can account for the production of the TeV
emission close to the core via inverse Compton scattering of
sub-mm disk photons by accelerated electrons. As it is unlikely that 
protons could be accelerated to EeV energies in this region, they
propose shear acceleration along the kpc-scale jet as the origin
for these particles.

\section{Conclusions}

Centaurus~A has been a popular potential source of cosmic rays
for close to half a century.  A number of cosmic ray and VHE
gamma-ray searches for excesses from that region have been made.
An early search by \citet{Grindlay1,Grindlay2} with VHE
gamma-rays was encouraging, showing evidence for a positive
observation at a time of a large X-ray flare, and, very recently,
H.E.S.S.\ has also provided evidence that Cen~A is a VHE
gamma-ray source.  The Buckland Park air shower array found a
signal with appropriate spectral characteristics which included
evidence for CMB absorption at sub-PeV energies.  Also, recently,
the Pierre Auger Observatory has shown evidence of a clustering
of UHE cosmic ray events around Cen~A although without evidence
for a different spectrum to that of other directions.  Taken with
observations which have produced upper limits and recognising
that none of these observations had well defined {\it a priori}
statistical analysis procedures, one cannot say with confidence
that Centaurus~A is a major source at UHE energies.
However, it is the nearest example to us of one of the few
classes of objects which have been identified as having a
structure possibly capable of accelerating particles to the
highest observed energies.  Acceleration of cosmic ray
particles to the highest measured energies in Cen~A would be at the limit
of parameters associated with the acceleration process.  A
reduction in the cosmic ray flux (from the general direction of
Cen~A) above about 60\,EeV from the power law at lower energies
could be a source effect rather than a GZK cut-off, which would
(and may well) otherwise apply to more distant sources.

For the future, we clearly require more data.  Further work at VHE 
(H.E.S.S.) energies to better define the source region and any
possible extended structure would be great progress.  An
extension of the gamma-ray spectrum upwards towards the CMB
absorption feature using the \v{C}erenkov technique, with better
angular uncertainty than Buckland Park, such as the TenTen
concept \citep{TenTen}, would be a major asset.  Also, solid
confirmation and understanding of the Pierre Auger ``Centaurus A
excess'' is urgently needed.


\end{document}